\documentclass[aps,pre,twocolumn,groupedaddress]{revtex4-1}

\usepackage{graphicx,amsmath,amsfonts}
\usepackage{epstopdf}

\usepackage{pdfpages}

\makeatletter
\AtBeginDocument{\let\LS@rot\@undefined}
\makeatother

\newcommand{\E}{\mathbb{E}}
\newcommand{\eps}{\varepsilon}

\newcommand{\Ord}{\mathcal{O}}

\newcommand{\dif}{\mathrm{d}}

\newcommand{\xxi}{{\rm{x}}}

\usepackage[colorinlistoftodos]{todonotes}

\newcounter{jwcomment}

\newcounter{gagcomment}

\begin{document}


\title{Stochastic model reduction for slow-fast systems with moderate time-scale separation}

\author{Jeroen Wouters}
\email{jwouters@nbi.ku.dk\\ Current affiliation: Niels Bohr Institute, University of Copenhagen, Copenhagen, Denmark}
\affiliation{Department of Mathematics and Statistics, University of Reading, Reading, UK}
\author{Georg A. Gottwald}
\email{georg.gottwald@sydney.edu.au}
\affiliation{School of Mathematics and Statistics, University of Sydney, NSW Sydney 2006, Australia.}

\date{\today}

\begin{abstract}
We propose a stochastic model reduction strategy for deterministic and stochastic slow-fast systems with finite time-scale separation. The stochastic model reduction relaxes the assumption of infinite time-scale separation of classical homogenization theory by incorporating deviations from this limit as described by an Edgeworth expansion.
A surrogate system is constructed the parameters of which are matched to produce the same Edgeworth expansions up to any desired order of the original multi-scale system. We corroborate our analytical findings by numerical examples, showing significant improvements to classical homogenized model reduction.
\end{abstract}


\maketitle

\section{Introduction}
Complex systems in nature and in the engineered world often exhibit a multi-scale character with slow variables driven by fast dynamics. For example, large proteins \cite{kamerlin_coarse-grained_2011} and the climate system \cite{PalmerWilliams} exhibit both fast, small scale fluctuations and slow, large scale transitions. The high complexity often puts the system out of reach of both analytical and numerical approaches. Typically one is only interested in the dynamics of the slow variables or observables thereof. It is then a formidable challenge to distill reduced slow equations which can make the problem amenable to theoretical analysis, allowing to identify relevant physical effects, or, from a computational perspective, allow for a larger computational time step tailored to the slow time scale. 
Homogenization theory \cite{GivonEtAl04,PavliotisStuart} derives reduced slow dynamics by assuming an infinitely large time-scale separation between slow and fast variables. It has been rigorously proven for multi-scale systems with stochastic \cite{Khasminsky66,Kurtz73,Papanicolaou76} and deterministic chaotic fast dynamics \cite{MelbourneStuart11,GottwaldMelbourne13,KellyMelbourne17} and has been applied with great success in the design of numerical algorithms for molecular dynamics \cite{EEngquist03,KevrekidisGearEtAl03} and in stochastic climate modelling \cite{MTV99,Majda03}. Several challenges remain, however, in formulating reliable stochastic slow limit systems. Whereas homogenization is rigorously proven only for the limiting case of infinite time scale separation, this assumption is never met in the real world. Hence, homogenized stochastic systems may fail in reproducing the statistical behaviour of the underlying deterministic multi-scale system for finite time-scale separation when an intricate interplay between the fast degrees and the slow degrees of freedom is at play.\\ 
Homogenization relies on the fact that the slow dynamics experiences the integrated effect of, in the limit of infinitely fast dynamics, infinitely many fast events. Therefore homogenization is in effect a manifestation of the central limit theorem (CLT). Finite time scale effects are then akin to finite sums of random variables. In the context of random variables, corrections to the CLT for sums of finite length $n$ can be described by the Edgeworth expansion, which provides an expansion of the distributions of sums, asymptotic in $1/\sqrt{n}$ \cite{BhattacharyaRanga}. Edgeworth expansions have been developed for independent and for weakly dependent identically distributed random variables \cite{GoetzeHipp94}, continuous-time diffusions \cite{ait-sahalia_maximum_2002} and ergodic Markov chains \cite{HervePene10}. We extend their applicability here to multi-scale systems, including the deterministic case. The Edgeworth expansion is universal in the sense that it is agnostic about the microscopic details of the fast process. Only integrals over higher-order correlation functions appear in the analytical expressions we obtain. We will use this aspect of Edgeworth expansions to derive a reduced model by constructing a low-dimensional surrogate model with the same Edgeworth corrections as the original multi-scale model. We show that this surrogate system is far superior to homogenization in reproducing the statistical behaviour of the slow dynamics.\\
The paper is organised as follows. In Section~\ref{s.model} we introduce the multi-scale systems under consideration and their diffusive limits in the case of infinite time scale separation as provided by homogenization theory. In Section~\ref{s.Edge} we establish corrections to the homogenized limit using Edgeworth expansions. These are then used in Section~\ref{s.surrogate} to construct a reduced surrogate stochastic model which captures finite time-scale separation effects.
We conclude in Section~\ref{s.discussion} with a discussion and an outlook.


\section{Multi-scale systems}
\label{s.model}
We consider multi-scale systems of the form
\begin{eqnarray}
\dif x & = &\frac{1}{\varepsilon} f_0 (x, y)\, \dif t + f_1(x,y)\, \dif t
\label{e.x}
\\
\dif y & = &\frac{1}{\varepsilon^2} g_0 (y)\, \dif t + \frac{1}{\varepsilon} \beta(y)\, \dif W_t  + \frac{1}{\varepsilon} g_1 (x, y)\, \dif t 
,
\label{e.y}
\end{eqnarray}
with slow variables $x\in \mathbb{R}^d$ and fast variables $y\in \mathbb{R}^N$. We assume that the fast dynamics $\dif y = g_0 \dif t + \beta \dif W_t$ admits a unique invariant physical measure $\nu$ and the full system admits a unique invariant physical measure $\mu^{(\eps)}(\dif x,\dif y)$ \footnote{An ergodic measure is called physical if for a set of initial conditions of nonzero Lebesgue measure the temporal average of a typical observable converges to the spatial average over this measure.}.  
The system may be stochastic with a non-zero diffusion matrix $\beta\in \mathbb{R}^{d\times l}$ and $l$-dimensional Brownian motion $\dif W_t$, or may be deterministic with $\beta\equiv 0$. In the latter case we assume that the fast dynamics is sufficiently chaotic \footnote{The assumptions on the chaoticity of the fast subsystem are mild. For continuous-time fast system, an associated Poincar\'e map needs to have a 
summable correlation function (irrespective of the mixing properties of the flow). Systems with such mild conditions on the chaoticity include, but go far beyond, Axiom A diffeomorphisms and flows, H\'enon-like attractors and Lorenz attractors; see \cite{MelbourneNicol05,MelbourneNicol08,MelbourneNicol09}}.\\ Homogenization theory deals with the limit of infinite time-scale separation $\eps\to 0$. In this limit it is well known that when the leading slow vector field averages to zero, i.e. $\langle f_0(x,y)\rangle = 0$, where 
$\langle A(y) \rangle := \int \nu(\dif y) A(y)$, the slow dynamics is approximated by an It\^o stochastic differential equation \cite{Khasminsky66,Kurtz73,Papanicolaou76,MelbourneStuart11,GottwaldMelbourne13c,KellyMelbourne14} of the form
\begin{equation}
d X = F(X) \dif t + \sigma(X) \, \dif W_t \;.
\label{e.homog}
\end{equation}
The drift coefficient is given by
\begin{eqnarray}
F(x) = & \;  \langle f_1(x,y)\rangle
  + \int_0^\infty \dif s \left( 
\langle f_0(x,y)\cdot \nabla_x f_0(x,\varphi^ty) \rangle 
\right.
\nonumber \\
& \hphantom{+ \int_0^\infty \dif s ()}
+ \left.
\langle g_1(x,y) \cdot \nabla_y \left(f_0(x,\varphi^ty)\right)\rangle \right),
\label{eq.contF} 
\end{eqnarray}
where $\varphi^t$ denotes the flow map of the fast dynamics, and the diffusion coefficient is given by the Green-Kubo formula
\begin{eqnarray}
\sigma(x) \sigma^T(x) &=&\int_0^\infty \dif s \left\langle  f_0(x,y) \otimes f_0(x,\varphi^ty) \right. \nonumber \\ && \qquad \qquad \left. + f_0(x,\varphi^ty) \otimes f_0(x,y) \right\rangle , 
\label{e.sGK}
\end{eqnarray}
where the outer product between two vectors is defined as $(a\otimes b)_{ij} = a_{i}b_{j}$ \footnote{As stated here the formulae for the drift and diffusion matrix are only valid for correlation functions which are slightly more than integrable. When the autocorrelation function of the fast driving system is decaying but is only integrable, more complicated formulae apply; see \cite{KellyMelbourne17} for details.}. For details the reader is referred to \cite{KellyMelbourne14}.


\section{Edgeworth approximation for dynamical systems}
\label{s.Edge}
There are three distinct time scales in the system \eqref{e.x}-\eqref{e.y}: a fast time scale of $\Ord(\eps^2)$, an intermediate time-scale of $\Ord(\eps)$ on which the fast dynamics has equilibrated but the slow dynamics has not yet evolved, and a long diffusive time scale of $\Ord(1)$ on which the slow variables exhibit non-trivial dynamics. It is on the intermediate time scale that we can expect corrections to the CLT: the time scale is sufficiently long for the fast dynamics to generate near-Gaussian noise but not long enough for the slow dynamics to dominate. This is reflected in the homogenized SDE (\ref{e.homog}): the slow variable is near-Gaussian with $dX \sim \sigma(X) \, \dif W_t$ exactly on the time scale $\sqrt{dt} \sim \eps$. We therefore focus our attention on the limit $\eps \rightarrow 0$ with $t/\eps=\theta$ constant, and study the transition probabilities between initial conditions $x_0$ into the interval $\chi = (\xxi, \xxi + \mathrm{d} \xxi)$
\begin{equation*}
\pi_\eps(\xxi,t,x_0) = \mathbb{P} \left( \left. \frac{x(t) - x_0 }{\sqrt{t}} \in \chi \right| x(0) = x_0, \, y(0) \sim   \mu_{x_0}^{(\eps)}\right) \,.
\end{equation*}
Here $\mu_{x_0}^{(\eps)}$ denotes the conditional measure of $\mu^{(\eps)}$ conditioned on $x=x_0$. 
In the limit of homogenization theory $\eps\to 0$, the transition probability $\pi_\eps$ converges to a normal distribution $\mathbf{n}_{0,\sigma^2}(\xxi)$ with the covariance given by the Green-Kubo formula (\ref{e.sGK}). For finite $\eps$, the transition probability will not be Gaussian but will have correction terms of $\mathcal{O}(\sqrt{\eps})$, the so called Edgeworth corrections. The corrections to the limiting Gaussian distribution of $\hat x(t) := (x(t) - x_0)/\sqrt{t}$ are most readily calculated through the characteristic function $\chi_\eps(\zeta)= \mathbb{E}_\eps^{x_0,\mu}\left[  \exp(i\zeta \hat x)\right]$ where $\E_\eps^{x_0, \mu}$ is the expectation value w.r.t. $\pi_\eps$.
Since $\ln \chi_\eps = \sum_n c_\eps^{(n)} (i \zeta)^n/n!$ with the cumulants of $\hat x$ 
\begin{eqnarray*}
c_\eps^{(p)} & = m_\eps^{(p)} - \sum_{j = 1}^{p - 1} 
{{p - 1}\choose{j - 1}}
m_\eps^{(p - j)} c_\eps^{(j)}\, ,
\end{eqnarray*}
and $m^{(p)}_\eps = \mathbb{E}^{x_0,\mu}_\eps [\hat x^p]$, we seek an asymptotic expansion
\begin{eqnarray*}
c_\eps^{(p)} = c_0^{(p)}+\sqrt{\eps} c_{\frac{1}{2}}^{(p)}+\eps c_1^{(p)} + {\mathcal{O}}(\eps^{\frac{3}{2}}) \,.
\end{eqnarray*}
To this end, the expectation values $\E_\eps^{x_0, \mu}$ are expressed as
\begin{eqnarray*}
\E_\eps^{x_0, \mu} \left[ A(x(t),y(t)) \right] 
& = \int \int A(x,y) e^{\mathcal{L}_\eps t} \delta_{x_0}(\dif x)\mu(\dif y) \, ,
\end{eqnarray*}
with the transfer operator $e^{\mathcal{L}_\eps t}$ (also known as Frobenius-Perron operator) associated with the multi-scale system (\ref{e.x})-(\ref{e.y}). 
This transfer operator can be expanded by successive application of the Duhamel-Dyson formula \cite{EvansMorris,Zwanzig}, resulting in explicit expressions for the $c^{(p)}_{j}$. We find $c_0^{(1)}=c_1^{(1)}=0$, $c_{\frac{1}{2}}^{(1)}=F(x_0)$, $c_0^{(2)}=\sigma^2$, $c_{\frac{1}{2}}^{(2)}=0$, $c_0^{(3)}=c_1^{(3)}=0$, $c_0^{(4)}=c_{\frac{1}{2}}^{(4)}=0$ and $c^{(p)}_\eps=\mathcal{O}(\eps^{\frac{3}{2}})$ for $p>4$. The coefficients $c_1^{(2)}$, $c_{\frac{1}{2}}^{(3)}$ and $c_{1}^{(4)}$ depend non-trivially on the correlations of $y$ (see \cite{WoutersGottwald17}).
%
Furthermore, by taking the inverse Fourier transform of $\chi_\eps$, we can expand the probability density $\pi_\eps = \pi^{(2)}_\eps + \mathcal{O}(\eps^{\frac{3}{2}})$ with
\begin{widetext}
\begin{eqnarray}
\pi^{(2)}_\eps(\xxi,t=\theta \eps,x_0) = \mathbf{n}_{0,\sigma^2} (\xxi) \,
\Bigg[
1 
&+ \sqrt{\eps} \left( \frac{F(x_0)}{\sigma} H_1 \left(\frac{\xxi}{\sigma}\right) 
+ \frac{c_{\frac{1}{2}}^{(3)}}{3! \sigma^3} H_3 \left(\frac{\xxi}{\sigma}\right) \right) 
+ \eps \left(\frac{ {F(x_0)}^2+c_1^{(2)}}{2 \sigma^2}  H_2 \left(\frac{\xxi}{\sigma}\right) 
\right.
\nonumber\\
&+ \frac{c_1^{(4)} + 4F(x_0) c_{\frac{1}{2}}^{(3)}}{4! \sigma^4}  H_4 \left(\frac{\xxi}{\sigma}\right) 
+ \frac{{c_{\frac{1}{2}}^{(3)}}^2}{2 (3! \sigma^3)^2} H_6 \left(\frac{\xxi}{\sigma}\right) 
\Big)        
\Bigg] 
+ \Ord(\eps^{\frac{3}{2}}) \,.
\label{e.gamma_xi}   
\end{eqnarray}
\end{widetext}
Here $H_n(\xxi) = (\xxi - \frac{\dif}{\dif \xxi})^n 1$ are Hermite polynomials of degree $n$. It is readily seen from (\ref{e.gamma_xi}) that for $\eps\to 0$, the homogenization limit $\lim_{\eps\to 0}\pi_\eps = \mathbf{n}_{0,\sigma^2}$ is recovered. For a rigorous derivation of the Edgeworth expansion and explicit formulae for the $c^{(p)}_j$ the reader is referred to \cite{WoutersGottwald17}. For completeness we have summarized the formulae in the Supplementary Material.\\  

\begin{figure}[h]
	\begin{tabular}{ll}
		\includegraphics[width = 0.85\columnwidth]{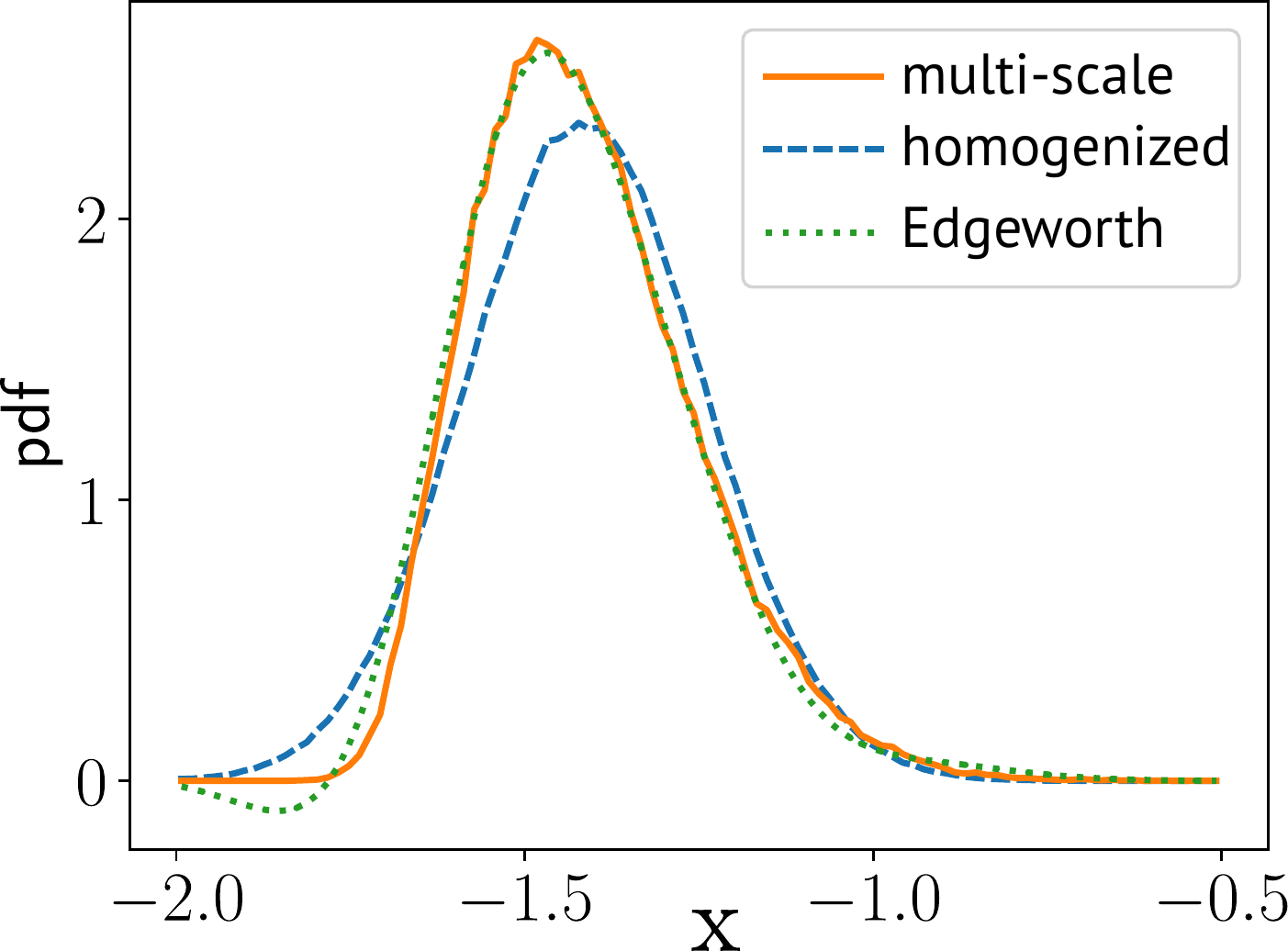}
	\end{tabular}
	\caption{Transition probability $\pi_\eps(\xxi,t=0.5,x_0=-\sqrt{2})$ of the system \eqref{eq.multiL96x}-\eqref{eq.multiL96y} (labelled ``multiscale'') with $a=0.2$, $b=0.1$, $\varepsilon=0.5$ and $\sigma_m=0.03642$ (implying $\sigma = 0.25$), the Edgeworth expansion \eqref{e.gamma_xi} (labelled ``Edgeworth'') and the pdf of $X(t)$ in \eqref{e.homog} (labelled ``homogenized'').}
	\label{fig.verifEdgew}
\end{figure}

We now demonstrate the validity of the Edgeworth expansion for the multi-scale system of the form \eqref{e.x}-\eqref{e.y}. In particular, we consider
\begin{eqnarray}
\dot{x} &=&  \frac{\sigma_m}{\eps} f_0(y) -\partial_x V(x) \label{eq.multiL96x} \\
\dot{y}_i &=& \frac{1}{\eps^2} \left( y_{i-1}(y_{i+1} - y_{i-2}) + R - y_i \right) \label{eq.multiL96y}
\end{eqnarray}
with $y_{N+i} = y_{i}$ for $1\le i \le N$. The system consists of a single degree of freedom $x$ in a symmetric double well potential $V(x)=x^2 (b^2 x^2 - a^2)$ driven by a fast Lorenz '96 system (L96). The L96 system was introduced to model the atmosphere in the midlatitudes \cite{Lorenz96}. The system \eqref{eq.multiL96x}-\eqref{eq.multiL96y} can be viewed as a simple toy model of the ocean exhibiting two regimes which is driven by a fast chaotic atmosphere. We take the classical parameters of Lorenz' with $N=40, \,R=8$ and choose  $f_0(y) = \frac{1}{5}\sum_{i=1}^{5} y_i^2 - C_0$ where $C_0$ is chosen such that $\langle f_0 \rangle=0$. Randomness is introduced solely through a random choice of the initial condition $y_0$, distributed according to the physical invariant measure of the fast L96 system. 
\\
To demonstrate the validity of the Edgeworth expansion we show in Figure~\ref{fig.verifEdgew} the transition probabilities for the full multi-scale system \eqref{eq.multiL96x}-\eqref{eq.multiL96y} as well as those of the reduced homogenized system \eqref{e.homog} and of the Edgeworth expansion \eqref{e.gamma_xi}.
Whereas homogenization fails to approximate the transition probability (with a relative error in the skewness of $0.87$), our Edgeworth approximation describes the statistics of the true system remarkably well.
See the Supplementary Material for more details on the numerical calculation of the Edgeworth coefficients.

\section{The surrogate system}
\label{s.surrogate}
The Edgeworth expansion is universal in the sense that only a limited number of statistical properties of the fast system appear in the expansion. Therefore, the microscopic details of the fast $y$-dynamics are of no importance to the slow $x$-dynamics. In fact, from the macroscopic point of view the $y$-dynamics can be substituted with a simpler surrogate system, as long as the statistical properties encoded in the Edgeworth expansion are preserved. This suggests a new way of performing stochastic model reduction for the slow dynamics: construct a class of simple surrogate systems $(X(t),\eta(t))$ with $X \in \mathbb{R}^d$ and $\eta \in \mathbb{R}^k$ with $k<N$, dependent on a set of parameters $\frak{p}_{\rm{surr}}$, the Edgeworth expansion of which can be analytically performed. Then judiciously choose the free parameters of the surrogate system $\frak{p}_{\rm{surr}}$ such that its Edgeworth corrections match the observed Edgeworth corrections of the original multi-scale model we set out to model.
Specifically, the transition probability of $X$,
\begin{equation*}
\pi_{\rm{surr}}(\xxi,t=\eps,x_0) = \mathbb{P} \left( \left. \frac{X(t) - X(0) }{\sqrt{t}} \in \chi \right| X(0) = x_0 \right),
\end{equation*}
is approximated by the second order Edgeworth expansion $\pi_{\rm{surr}} = \pi^{(2)}_{\rm{surr}} + \mathcal{O}(\eps^{\frac{3}{2}})$ and 
the parameters by the constrained optimization
\begin{eqnarray}
{\arg\hspace{-0.0cm}\min_{\hspace{-0.4cm}\frak{p}_{\rm{surr}}}\hphantom{\,} } \left\| \pi^{(2)}_{\rm{surr}}(\xxi,\eps,x_0) - \pi^{(2)}_\eps(\xxi,\eps,x_0)  \right\|
\label{e.argmin}
\end{eqnarray}
of the $L_2$-norm subject to the exact matching of the leading order diffusivity (\ref{e.sGK}) and drift (\ref{eq.contF}). These constraints assure that the surrogate system and the full deterministic system have the same homogenized limiting equation. 

We consider here the following family of surrogate models for the multi-scale system (\ref{e.x})-(\ref{e.y}) 
\begin{eqnarray}
\dot X &=& \frac{1}{\eps} f_0^{(s)}(X,\eta) + F(X) + f_1^{(s)}(X,\eta)
\label{e.surr0x}
\\
d\eta &=& -\frac{1}{\eps^2} \Gamma^{(s)} \eta \, \dif t +\frac{\sigma^{(s)}}{{\eps}} \dif W_t + \frac{1}{\eps} g_1^{(s)}(X,\eta)\, .
\label{e.surr0y}
\end{eqnarray}
The fast process $\eta(t)$ is a $k$-dimensional Ornstein-Uhlenbeck process with $\Gamma^{(s)}_{ij}=\gamma_i \delta_{ij}$ and $\sigma^{(s)}_{ij}=\zeta_i\delta_{ij}$. The noise is here, different to the homogenized diffusive limits, coloured and enters the slow dynamics in an integrated way, allowing for non-trivial memory.\\ The vector fields $f_0^{(s)}$, $f_1^{(s)}$ and $g_1^{(s)}$ of the surrogate system are chosen to be polynomial
\begin{eqnarray}
f_l^{(s)}(X,\eta) 
&= 
\sum_{|\alpha|<\alpha_l, |\beta|<\beta_l}
a_l^{(\alpha,\beta)}X^\alpha\, \eta^\beta\\
g_1^{(s)}(X,\eta) 
&= 
\sum_{|\alpha|<\alpha_2, |\beta|<\beta_2}
a_2^{(\alpha,\beta)}X^\alpha\, \eta^\beta
\, 
\end{eqnarray}
for $l=0,1$. The degree of the polynomials $\alpha_l$ and $\beta_l$, $l=0,1,2$, and the dimensionality of the surrogate process $k$ are chosen as the smallest degree and dimension which still allows the surrogate system to capture the statistical features of the vector field $f_0(x,y)$ of the original multi-scale system \eqref{e.x}-\eqref{e.y}, such as the support of its empirical measure. For the application considered here, we find that $k=1$, $\alpha_{1,2}=\beta_{1,2}=0$, $\alpha_0=3$, $\beta_0=1$ are sufficient. Analytical expressions for the Edgeworth coefficients for this system are reported in the Supplementary Material.
The parameter $a_0^{(0,0)}=-a_0^{(0,2)} \zeta_1^2 / (2 \gamma_1)$ is determined by requiring the centering condition $\langle f_0^{(s)}\rangle\equiv 0$.


%

\begin{figure}[h]
  \begin{tabular}{ll}
     \includegraphics[width = 0.85\columnwidth]{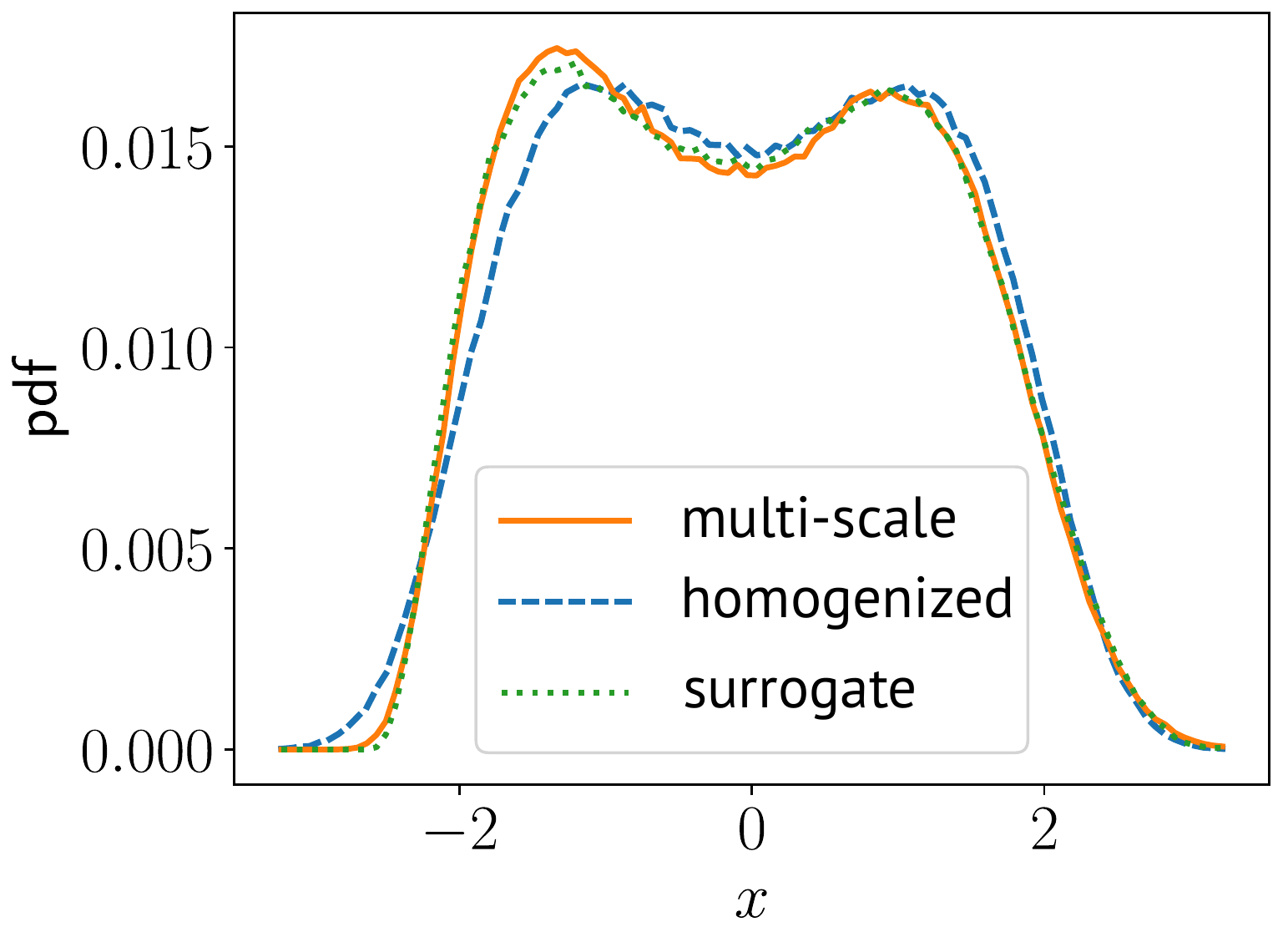}\\
     \includegraphics[width = 0.85\columnwidth]{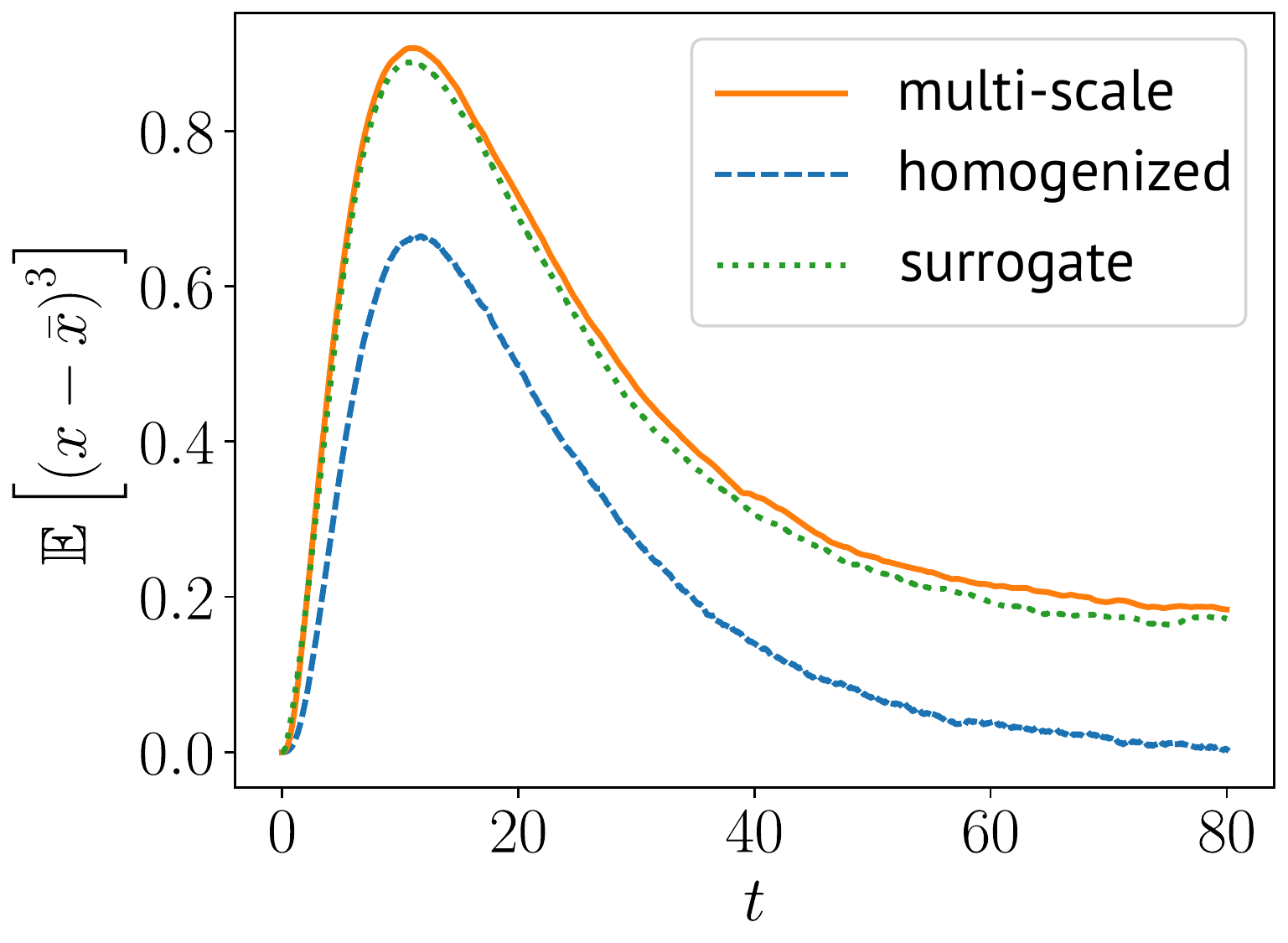}
  \end{tabular}
  \caption{The invariant measure (top) and third moment (bottom) for $x$ of the multi-scale Lorenz system \eqref{eq.multiL96x} with  $\varepsilon=1$, $a=0.15$, $b=0.1$ and $\sigma_m=0.07285$ (implying $\sigma = 0.5$), the homogenized equation \eqref{e.homog} and the surrogate process \eqref{e.surr0x}-\eqref{e.surr0y}.  The parameters of the surrogate process are obtained by the method in the main text as $\gamma_1=2.479$, $\zeta_1=25.793$, $a_0^{(0,3)} =-1.462 \; 10^{-3}$, $a_0^{(0,2)}=2.958 \; 10^{-2}$,  $a_0^{(0,1)}=1.079$.}

\label{fig.surrogate}
\end{figure}

Figure \ref{fig.surrogate} shows the invariant measure and the third moment of the slow dynamics of the multiscale Lorenz system \eqref{eq.multiL96x} with a moderate time scale separation $\eps=1$, as well as of the homogenized  equation \eqref{e.homog} and of the surrogate process \eqref{e.surr0x}-\eqref{e.surr0y}. It is clearly seen that the stochastic model reduction based on the Edgeworth expansion captures the nontrivial non-Gaussian behaviour of the full slow dynamics very well, whereas the homogenized equation converges to a Gaussian with a zero third moment. In the Supplementary Material we provide numerical results for the first four moments, highlighting the accuracy of the surrogate model in capturing the statistical behaviour of the slow dynamics for moderate time scale separation.

\section{Discussion}
\label{s.discussion}

We developed a new framework in which to perform stochastic model reduction of multi-scale systems with moderate time scale separation. We showed how Edgeworth expansions can be used to construct reduced models for the slow dynamics of a chaotic deterministic multi-scale model. 
We considered a family of surrogate models where the free parameters were chosen to match the Edgeworth expansion of the original multi-scale model under consideration. The degree of the surrogate model was chosen by assuring to have the lowest possible order of the polynomials while still allowing for the surrogate system to capture the overall statistical features of the full multi-scale system. Matching the Edgeworth expansion then singles out the optimal member in the prescribed class. We remark that the Edgeworth expansion is based on the transition probability on the intermediate time scale.
In some applications, such as weather forecasting, one is interested in the transitional dynamics and their statistical modelling rather than in the long term statistical behaviour. In this situation Edgeworth expansions allow for a faithful description of the effects of finite time scale separation.
The aim of the reduced model in other applications, however, may be to describe the statistical behaviour on the longer diffusive time scale, for example in climate science. We observe that in the system considered here, matching the short time transition probabilities translates into a more reliable description of the long time statistics as well. Although this property may not hold in general, we expect it to hold in sufficiently smooth systems. 
\\
Our framework is not limited to deterministic continuous time systems. It can be extended to stochastic multi-scale systems and to discrete time maps which would allow the study of numerical integrators and their statistical limiting behaviour of resolved modes. More importantly, Edgeworth approximations can be determined from observational data; this allows for the application to systems with high complexity prohibiting an analytical estimation of the Edgeworth corrections. This opens up the door to perform mathematically sound stochastic model reductions for real-world problems. Furthermore, Edgeworth approximations are not limited to multi-scale systems. As an extension of the CLT, they can be used to study finite size effects to the thermodynamic limit of weakly coupled systems such as Kac-Zwanzig heat baths for distinguished particles \cite{FordEtAl65,Zwanzig73,FordKac87}.


\begin{acknowledgments}
The research leading to these results has received funding from the European Community's Seventh Framework Programme (FP7/2007-2013) under grant agreement n° PIOF-GA-2013-626210. We thank Ben Goldys and Fran\c{c}oise P\`{e}ne for enlightening discussions and comments.
\end{acknowledgments}


\bibliography{Edgeworth}

\newpage\newpage


\section*{Supplementary material}

# Stochastic model reduction for slow-fast systems with moderate time-scale separation
# Supplementary material

Jeroen Wouters[*]

*Department of Mathematics and Statistics,*

*University of Reading, Reading, UK*

Georg A. Gottwald[†]

*School of Mathematics and Statistics,*

*University of Sydney, NSW Sydney 2006, Australia.*

(Dated: April 10, 2018)

**EXPLICIT EXPRESSION OF THE EDGEWORTH COEFFICIENTS**

In order to employ the Edgeworth expansion or to construct the surrogate model, we first need to determine the Edgeworth coefficients $\sigma$, $c_1^{(2)}$, $c_{\frac{1}{2}}^{(3)}$ and $c_1^{(4)}$ appearing in Equation (6) of the main text.

For the case of the multi-scale Lorenz '96 system in Eqs. (7)-(8) of the main text, these are given by [1]:

$$F = -\partial_{x_0} V(x_0)$$

$$\sigma^2 = \sigma_m^2 \, \mu_{20}$$

$$c_1^{(2)} = \frac{\sigma_m^2}{\theta^2} \mu_{21} - \sigma^2 \partial_{x_0}^2 V(x_0)$$

$$c_{\frac{1}{2}}^{(3)} = \frac{\sigma_m^3}{\theta} \mu_{30}$$

$$c_1^{(4)} = \frac{\sigma_m^4}{\theta^2} \mu_{40}$$

where

$$\mu_{20} = 2 \int_0^\infty C_2(\tau) \, \mathrm{d}\tau \tag{1}$$

$$\mu_{21} = -2 \int_0^\infty \tau \, C_2(\tau) \, \mathrm{d}\tau \tag{2}$$

$$\mu_{30} = 6 \int_0^\infty C_3(\tau_1, \tau_2) \, \mathrm{d}\tau_1 \mathrm{d}\tau_2 \tag{3}$$

$$\mu_{40} = 6\mu_{20}\,\mu_{21} - 24 \int_0^\infty \left( C_4(\tau_1, \tau_2, \tau_3) - C_2(\tau_1)C_2(\tau_3) \right) \, \mathrm{d}\tau_1 \mathrm{d}\tau_2 \mathrm{d}\tau_3 \tag{4}$$

with the two-point autocorrelation function $C_2(\tau) = \langle f_0(x) f_0(\varphi^\tau x) \rangle$, the three–point autocorrelation function $C_3(\tau_1, \tau_2) = \langle f_0(x) f_0(\varphi_1^\tau x) f_0(\varphi^{\tau_1+\tau_2} x) \rangle$ and the four–point autocorrelation function $C_4(\tau_1, \tau_2, \tau_3) = \langle f_0(x) f_0(\varphi^{\tau_1} x) f_0(\varphi^{\tau_1+\tau_2} x) f_0(\varphi^{\tau_1+\tau_2+\tau_3} x) \rangle$, where we recall that $\varphi^t$ denotes the flow map of the fast dynamics.

**NUMERICAL ESTIMATION OF THE EDGEWORTH COEFFICIENTS**

The terms $\mu_{20}$, $\mu_{21}$, $\mu_{30}$ and $\mu_{40}$ can be calculated directly by estimating the correlation functions $C_{2,3,4}$. This, however, is computationally very expensive to get accurate converged results. Here we estimate the terms as follows. As shown in [1], the Edgeworth coefficients

appear as the coefficients of an expansion in $t$ and $\varepsilon$ of the cumulants of transition probabilities of the multi-scale system. For $V = 0$, the terms $\mu_{20}$, $\mu_{21}$, $\mu_{30}$ and $\mu_{40}$ are the leading order terms appearing in the Edgeworth expansion of the second, third and fourth cumulant. More specifically, for the system

$$\dot{x} = \frac{1}{\varepsilon} f_0(y)$$
$$\dot{y} = \frac{1}{\varepsilon^2} g_0(y)$$

with initial conditions $x(t = 0) = x_0$ and $y(t = 0) = y_0$, we can integrate the slow dynamics to obtain

$$\zeta_\varepsilon := \frac{x(t = \varepsilon) - x_0}{\sqrt{\varepsilon}} = \sqrt{\varepsilon} z(\frac{1}{\varepsilon})$$

with $z(t) := \int_0^t f_0(y(\tau)) \, d\tau$. As shown in [1], the second, third and fourth cumulants of $\zeta_\varepsilon$ can be expanded in orders of $\sqrt{\varepsilon}$ as

$$\mathbb{E}_\varepsilon^{x_0,\mu}\left[\zeta_\varepsilon^2\right] = \mu_{20} + \varepsilon \mu_{21} + \mathcal{O}(\varepsilon^2)$$
$$\mathbb{E}_\varepsilon^{x_0,\mu}\left[\zeta_\varepsilon^3\right] = \sqrt{\varepsilon} \mu_{30} + \mathcal{O}(\varepsilon^{\frac{3}{2}})$$
$$\mathbb{E}_\varepsilon^{x_0,\mu}\left[\zeta_\varepsilon^4\right] - 3\mathbb{E}_\varepsilon^{x_0,\mu}\left[\zeta_\varepsilon^2\right]^2 = \varepsilon \mu_{40} + \mathcal{O}(\varepsilon^2).$$

It follows by taking $t = \frac{1}{\varepsilon}$ that $\mu_2 := \mathbb{E}\left[z(t)^2\right]$, $\mu_3 := \mathbb{E}\left[z(t)^3\right]$ and $\mu_4 := \mathbb{E}\left[z(t)^4\right]$ scale with $t$ as

$$\frac{\mu_2}{t} = \mu_{20} + \frac{\mu_{21}}{t} + \mathcal{O}(\frac{1}{t^2})$$
$$\frac{\mu_3}{t} = \mu_{30} + \mathcal{O}(\frac{1}{t})$$
$$\frac{\mu_4 - 3\mu_2^2}{t} = \mu_{40} + \mathcal{O}(\frac{1}{t}).$$

This suggests to perform a least-squares fit of $\frac{\mu_2}{t}$, $\frac{\mu_3}{t}$ and $\frac{\mu_4 - 3\mu_2^2}{t}$ to a two-parameter family of functions $\ell(t) = a + b/t$. Denoting the result of the least-square fit of $\frac{\mu_2}{t}$ by $a_2^\star$ and $b_2^\star$, of $\frac{\mu_3}{t}$ by $a_3^\star$ and $b_3^\star$ and of $\frac{\mu_4 - 3\mu_2^2}{t}$ by $a_4^\star$ and $b_4^\star$, we can extract the leading order coefficients. From the fits we obtain $\mu_{20} = a_2^\star$ and $\mu_{21} = b_2^\star$, $\mu_{30} = a_3^\star$ and $\mu_{40} = a_4^\star$.

Figure 1 shows the scaled cumulants of $z(t)$ together with their respective least squared fit of functions $\ell(t) = a + b/t$.

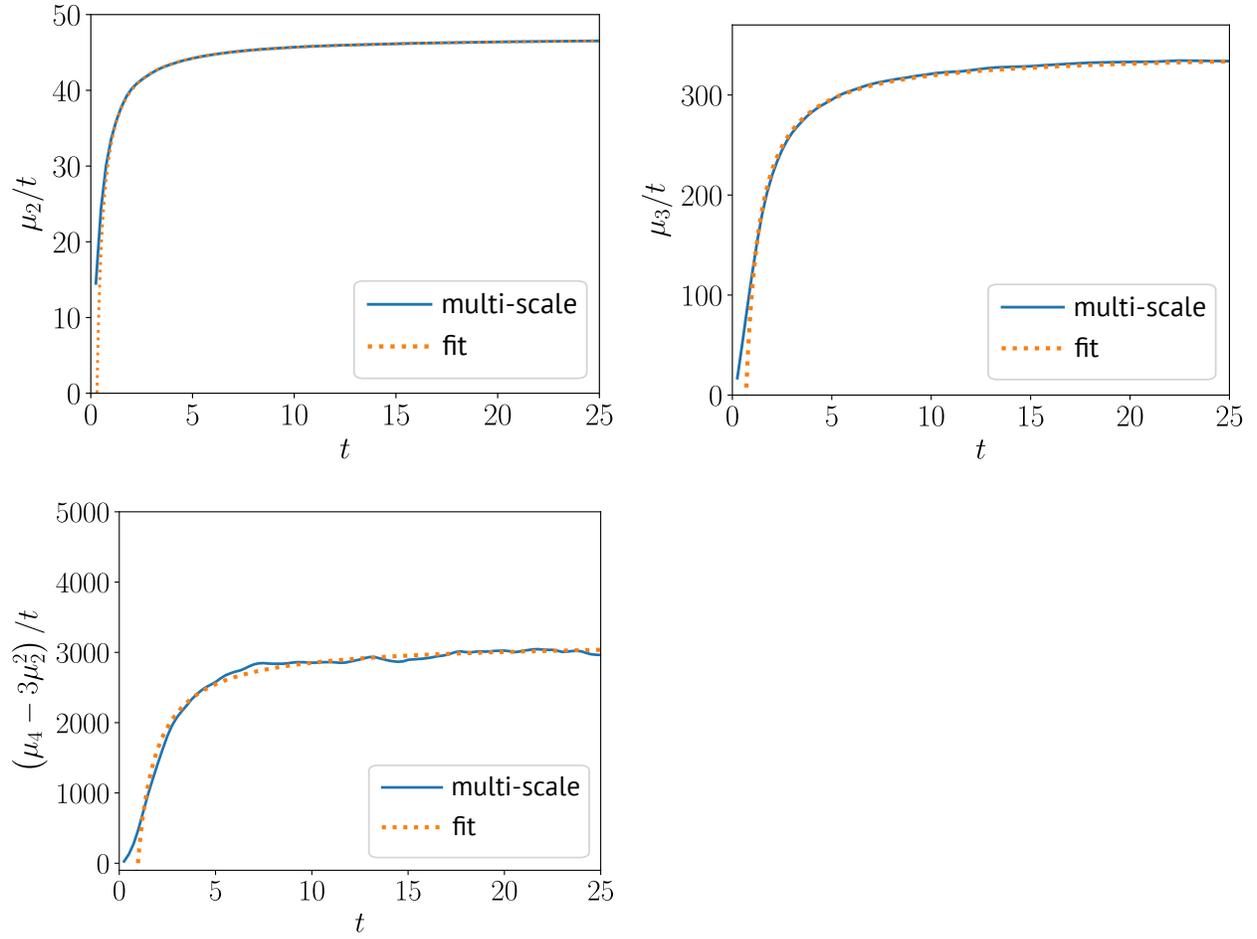

Figure 1: Scaled cumulants of $z(t)$ for the the full multi-scale system (Eqn (7) in main text) with $V \equiv 0$. The smooth line represents a least square fit to $\ell(t) = a + b/t$. Top left: second cumulant, top right: third cumulant, bottom: fourth cumulant.

# EDGEWORTH COEFFICIENTS FOR THE SURROGATE MODEL

For the surrogate system (10)-(11) of the main text the Edgeworth corrections (1)-(4) can be explicitly calculated. For $k=1$, $\alpha_{1,2} = \beta_{1,2} = 0$ and $\alpha_0 = 3$, $\beta_0 = 1$ we obtain

$$\mu_{20}^{(s)} = \frac{11\, a_0^{(3,0)^2}\zeta_1^6 + 4\, a_0^{(1,0)^2}\gamma_1^2\zeta_1^2 + 2\left(a_0^{(2,0)^2} + 6\, a_0^{(1,0)} a_0^{(3,0)}\right)\gamma_1\zeta_1^4}{4\,\gamma_1^4}$$

$$\mu_{21}^{(s)} = -\frac{29\, a_0^{(3,0)^2}\zeta_1^6 + 12\, a_0^{(1,0)^2}\gamma_1^2\zeta_1^2 + 3\left(a_0^{(2,0)^2} + 12\, a_0^{(1,0)} a_0^{(3,0)}\right)\gamma_1\zeta_1^4}{12\,\gamma_1^5}$$

$$\mu_{30}^{(s)} = \frac{3\left(22\, a_0^{(2,0)} a_0^{(3,0)^2}\zeta_1^8 + 4\, a_0^{(1,0)^2} a_0^{(2,0)}\gamma_1^2\zeta_1^4 + \left(a_0^{(2,0)^3} + 18\, a_0^{(1,0)} a_0^{(2,0)} a_0^{(3,0)}\right)\gamma_1\zeta_1^6\right)}{2\,\gamma_1^6}$$

$$\begin{aligned}\mu_{40}^{(s)} = {}& 6\,\mu_{20}^{(s)}\mu_{21}^{(s)} + \Big(48\,\gamma_1^4 a_1^4 + 420\,\gamma_1^3\zeta_1^2 a_1^2 a_2^2 + 66\,\gamma_1^2\zeta_1^4 a_2^4 + 480\,\gamma_1^3\zeta_1^2 a_1^3 a_3 \\ & + 2268\,\gamma_1^2\zeta_1^4 a_1 a_2^2 a_3 + 1976\,\gamma_1^2\zeta_1^4 a_1^2 a_3^2 + 3259\,\gamma_1\zeta_1^6 a_2^2 a_3^2 \\ & + 3912\,\gamma_1\zeta_1^6 a_1 a_3^3 + 3109\,\zeta_1^8 a_3^4\Big)\,\zeta_1^4/(8\,\gamma_1^9)\,.\end{aligned}$$

# FURTHER NUMERICAL RESULTS

In addition to the results for the invariant measure and third moment of the surrogate system presented in Fig. 2 in the main text, we present in Figure 2 here the second and fourth cumulants. We show results for the full multi-scale system (Eqn (7) in main text), the homogenized equation (Eqn (3) in main text) and the surrogate process (Eqns (10)-(11) in main text). For the second moment we show the long-time behaviour where homogenization matches well, as well as the intermediate time evolution where the Edgeworth expansion clearly outperforms the homogenized result. For the fourth moment the classical homogenization results fail to capture the long-time and the intermediate time temporal evolution whereas the Edgeworth expansion closely follows the true evolution of the moments, capturing the non-Gaussian behaviour of the slow dynamics in the moderate timescale separation case.

---

* j.wouters@reading.ac.uk

† georg.gottwald@sydney.edu.au

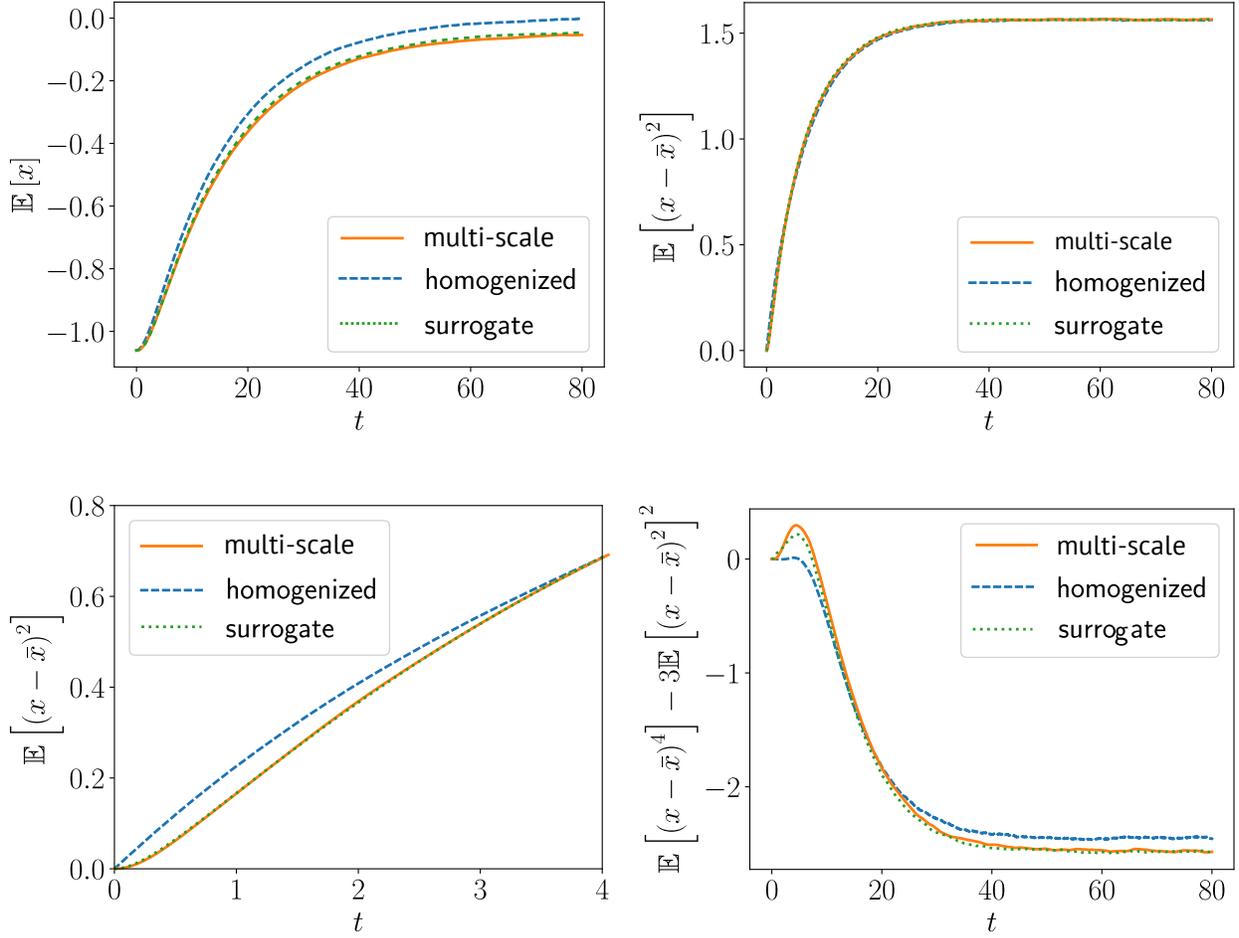

Figure 2: The first moment (top left), second moment (over long times (top right) and over intermediate times (bottom left)) and fourth cumulant over long times (bottom right) for $x$ of the multi-scale Lorenz system as a function of time. Parameters are $\varepsilon = 1$, $a = 0.15$, $b = 0.1$ and $\sigma_m = 0.07285$ (implying $\sigma = 0.5$). We show results for the full multi-scale system (Eqn (7) in main text), the homogenized equation (Eqn (3) in main text) and the surrogate process (Eqns (10)-(11) in main text). The parameters of the surrogate process are obtained by the method in the main text as $\gamma_1 = 2.479$, $\zeta_1 = 25.793$, $a_0^{(0,3)} = -1.462\ 10^{-3}$, $a_0^{(0,2)} = 2.958\ 10^{-2}$, $a_0^{(0,1)} = 1.079$ and $a_0^{(0,0)} = -a_0^{(0,2)}\zeta_1^2/(2\gamma_1)$.


\end{document}